%% file: main.tex
\documentclass[pdftex,twocolumn,epjc3,a4paper]{svjour3}

\usepackage[switch]{lineno}

\RequirePackage[T1]{fontenc}
\smartqed  
\usepackage{amsmath}
\usepackage{amssymb}
\RequirePackage{flushend}
\RequirePackage{mathrsfs}
\RequirePackage{dsfont}
\usepackage{ulem}
\usepackage{graphicx}
\usepackage[colorlinks = true,
            linkcolor = blue,
            urlcolor  = blue,
            citecolor = blue,
            anchorcolor = blue]{hyperref}
\usepackage{cite}
\usepackage[utf8]{inputenc}
\usepackage{xcolor} 
\usepackage{upgreek}
\usepackage[switch]{lineno}
\usepackage{caption}
\usepackage{subcaption}

\RequirePackage{mathptmx}      
\usepackage{blindtext}
\usepackage{units}
\usepackage{gensymb}
\journalname{.}

\begin{document}

\input{authors_cresst_EPJ_format}

\title{DoubleTES detectors to investigate the CRESST low energy background: results from above-ground prototypes}

\date{Received: date / Accepted: date}

\maketitle
\sloppy
\begin{abstract}
In recent times, the sensitivity of low-mass direct dark matter searches has been limited by unknown low energy backgrounds close to the energy threshold of the experiments known as the low energy excess (LEE). The CRESST experiment utilises advanced cryogenic detectors constructed with different types of crystals equipped with Transition Edge Sensors (TESs) to measure signals of nuclear recoils induced by the scattering of dark matter particles in the detector. In CRESST, this low energy background manifests itself as a steeply rising population of events below 200 eV. \\
A novel detector design named doubleTES using two identical TESs on the target crystal was studied to investigate the hypothesis that the events are sensor-related. We present the first results from two such modules, demonstrating their ability to differentiate between events originating from the crystal's bulk and those occurring in the sensor or in its close proximity. 
\end{abstract}

\section{Introduction}
Dark matter (DM) is one of the most investigated mysteries in modern physics, and unveiling its nature has motivated many experiments in the past decades. Direct detection experiments aim at measuring a DM particle scattering off a target nucleus. Despite the large effort of the scientific community in this direction, an unambiguous signal has not been found yet.\\ 
Recently, interest has grown for light DM candidates with masses in the sub-GeV/c$^2$ range.
The Cryogenic Rare Event Search with Superconducting Thermometers (CRESST) experiment is a direct dark matter detection experiment located in the underground Laboratori Nazionali del Gran Sasso (LNGS) in Italy. The CRESST cryogenic technology has proven to be one of the most favourable to explore this part of the parameter space \cite{APPECreport}, and currently provides the best sensitivity to the sub-GeV/c$^2$ mass range under standard assumptions\cite{Abdelhameed:2019hmk}. \\
CRESST operates cryogenic calorimeters consisting of a target crystal (mostly scintillating material) instrumented with a tungsten Transition Edge Sensor (W-TES) and paired with an auxiliary crystal (typically a thin silicon-on-sapphire wafer) also instrumented with a W-TES. This auxiliary detector is employed to measure the scintillation light emitted by the target crystal, hence named Light Detector (LD).
These detectors guarantee a low energy threshold of the order of tens of eV for nuclear recoils\cite{CRESST:2023cxk}, along with an excellent energy resolution. The employment of scintillating crystals enables effective particle discrimination through the simultaneous detection of light and phonon signals \cite{PhononLight, CRESST730kg}.  \\

Recent direct searches for light DM have encountered a challenge due to an unknown background at low energy, which limits their sensitivity \cite{osti_1862233}. In the CRESST experiment, this background manifests as a notable increase in detected events for energies below 200$\,$eV, with the event rate increasing with decreasing energy\cite{LEEdescr}. The exact cause of this phenomenon remains unclear.\\
Numerous other experiments focusing on low-mass DM searches reported the observation of an unknown background rising close to their energy threshold, including SuperCDMS \cite{CDMS}, DAMIC \cite{DAMIC}, EDELWEISS \cite{EDELWEISS}, NewsG \cite{NEWSG}, and SENSEI \cite{SENSEI}. Similar observations have been made in fields outside DM research, such as coherent neutrino-nucleus scattering (CE$\nu$NS) experiments like NUCLEUS \cite{NUCLEUS}, Ricochet \cite{Ricochet}, and MINER \cite{MINER}, among others. The LEE has attracted considerable attention within the scientific community (see \cite{osti_1862233} for an overview), though it is important to note that, while these observations share common features, the underlying causes may differ across experiments.\\
The LEE is characterised by a featureless spectrum, which could potentially be mistaken for a DM spectrum. However, the variations in the spectra observed dismiss this possibility. The CRESST collaboration initiated a comprehensive experimental campaign to explore the LEE's origin, detailed in \cite{LEEdescr} and \cite{LTD20}.  
Tests included comprehensive studies with different types of crystals, detector housing, and holding schemes to investigate potential sources of the LEE.\\
We recorded data continuously and performed offline triggering after filtering the stream with an optimum filter. The low thresholds achievable with offline triggering allow for the exploration of the LEE spectrum. 
To exclude the contribution of misidentified noise triggers to the LEE, we studied the spectra extracted from the inverted data stream. These tests showed that the small events constituting the LEE occur only with a positive trigger, while noise triggers would be expected with both polarities \cite{siliconpapero}. The observed events are, therefore, particle-like events. A LEE rate's decay pattern, with two distinct time constants—a short one over $\sim$15$\,$days and a longer one around 150$\,$days \cite{LEEdescr}—was also observed. The faster component was found to reactivate after thermal cycles from the working temperature of a few millikelvin to several tens of kelvin \cite{LEEdescr}.\\
These investigations have helped to dismiss several hypotheses, furthering our understanding of the LEE, though its precise origin remains unidentified. Some of the hypotheses still under consideration involve sensor-related events, such as events happening at the interface between the TES and the absorber or in the sensor itself. To investigate these hypotheses, the CRESST collaboration has developed the doubleTES detector design \cite{LTD20}.
This design is presented in Section \ref{sec:doubleTES_motivation}, and it has been produced on different absorber crystals. In the following Sections, we will present the results from above-ground tests obtained with CaWO$_\text{4}$ (Section \ref{CaWO4}) and silicon-on-sapphire substrates (Section \ref{SOS}).

\section{DoubleTES detectors}\label{sec:doubleTES_motivation}

The doubleTES detectors have two identical sensors on the absorber, each independently read out. These two W-TESs are fabricated onto the crystal surface in the same production process, ensuring highly uniform performance. This is a critical aspect of fabricating doubleTES detectors, as we will detail in Section \ref{CaWO4_results}. \\
In a W-TES the transition between the normal conducting phase and the superconducting phase of a tungsten film is used as a sensitive thermometer. The stabilisation of the sensor in the superconducting transition is achieved using a resistive heater. In the standard CRESST-III design, the heater is deposited near the W-TES on the crystal surface. In the doubleTES design, two distinct heaters are employed, each dedicated to stabilising one of the TESs. To reduce thermal interference between the sensors, these heaters are placed directly on the W-TESs and are electrically insulated from them by a 350$\,$nm thick SiO$_2$ layer.
\begin{figure}[b] 
\begin{center}
\includegraphics[width=1.0\linewidth, keepaspectratio]{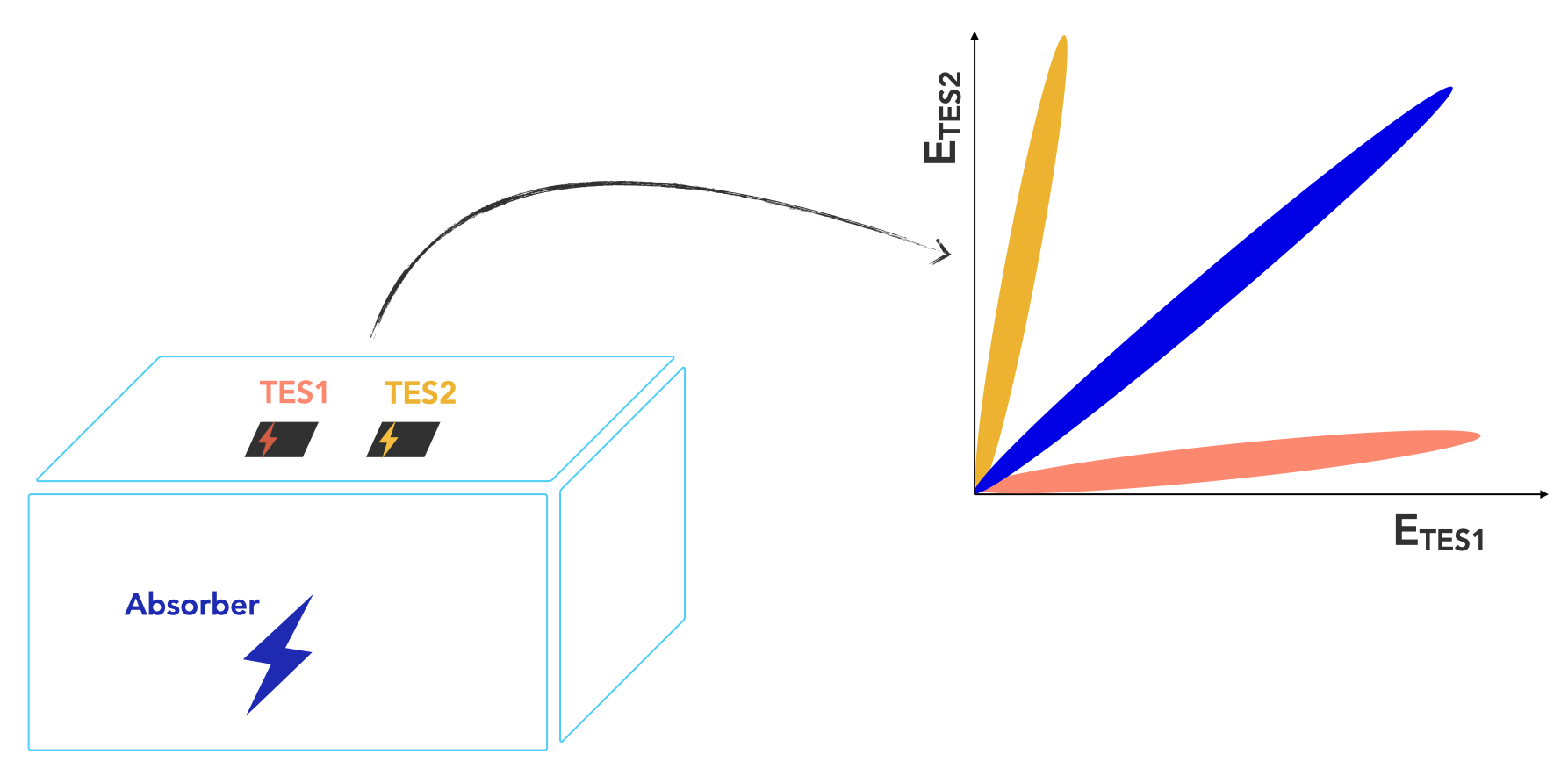}
\caption{Schematic representation of the signal originating from different sources in the doubleTES detector module. In a plot with the energies detected by the two TESs on the axes, we expect a clear separation of the events originating in the bulk of the crystal from those originating in close vicinity to one of the sensors.}\label{doubleTES_scheme}
\end{center}
\end{figure}
These detectors sit in a copper holder without any external force applied but their own weight. See \cite{LTD20} for a detailed description. \\
By analysing the data obtained from the two sensors independently, the doubleTES detectors can differentiate between events occurring within the bulk of the absorber and those in or near the sensors. The energy measured by both TESs should be similar for particle interactions in the crystal's bulk. Events originating in close proximity to one of the TESs are instead expected to have a stronger signal in one of the sensors (Figure \ref{doubleTES_scheme}).
This ability to discriminate is crucial for testing the hypothesis that the LEE originates from the interface between the crystal and sensor or from the sensor itself.\\

\section{CaWO$_4$ doubleTES detector}
\label{CaWO4}

\begin{figure}[b]
\begin{center}
\includegraphics[width=0.65\linewidth, keepaspectratio]{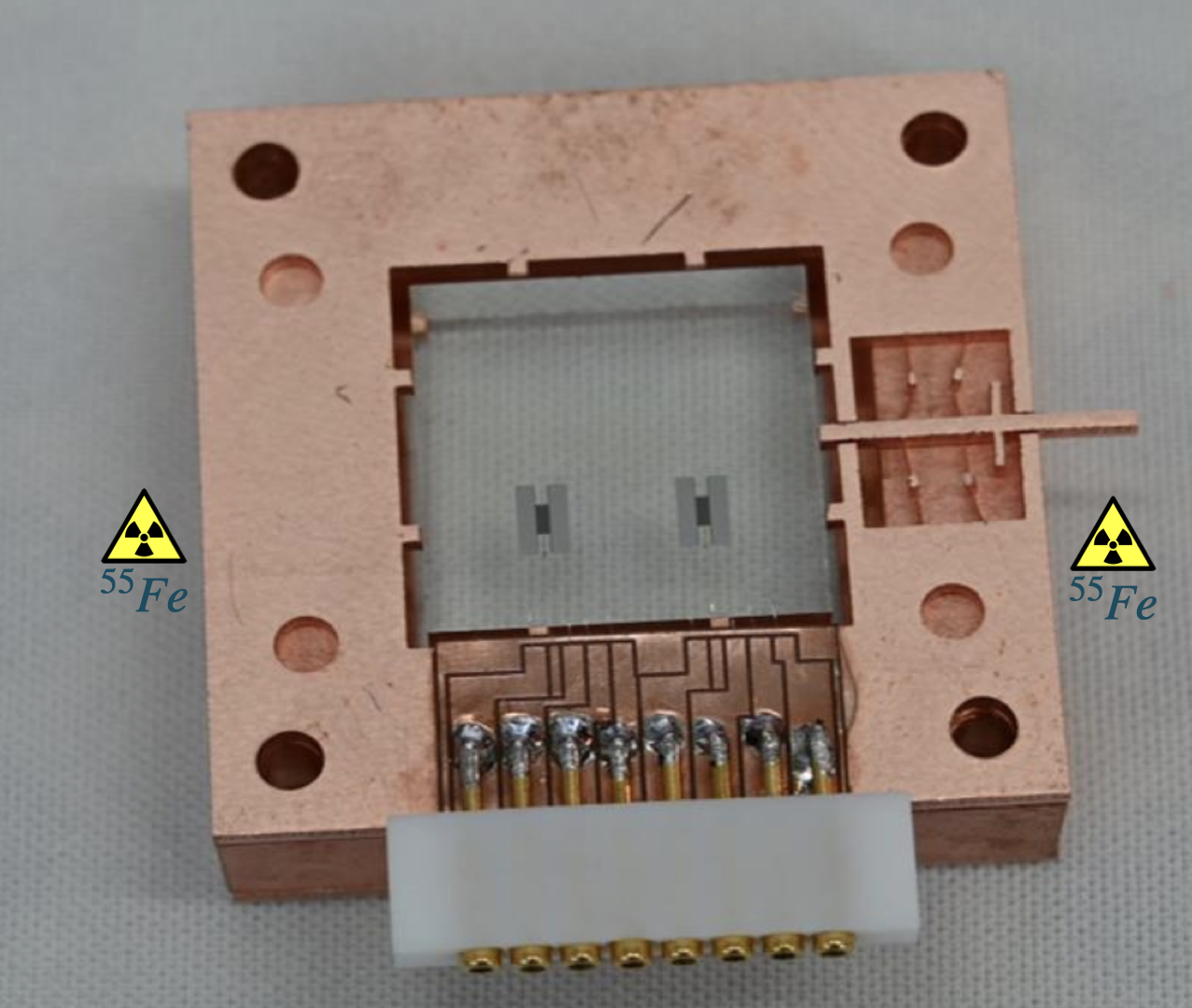}
\caption{A picture of the doubleTES CaWO$_4$ crystal, showing the two identical TESs seen from the top, mounted in the stress-free holding scheme. The position of the two $^{\text{55}}$Fe sources is indicated by the symbols.}\label{double-picture}
\end{center}
\end{figure}

The goal of this measurement is twofold: firstly, to demonstrate the operability of this module, and secondly, to verify its effectiveness in distinguishing events that originate in the TES. A picture of the detector, as tested in above-ground conditions, is provided in Figure~\ref{double-picture}.

\subsubsection*{Test results}
\label{CaWO4_results}

The CaWO$_4$ doubleTES detector has been the first prototype of this novel detector design. It consists of a 20$\times$20$\times$10$\,$mm$^3$ CaWO$_4$ crystal. It has been tested in the above-ground facility of the Max-Planck Institute for Physics in Munich, consisting of a dilution refrigerator surrounded by a 10$\,$cm thick lead wall to shield the experimental volume from external radiation.  
The detector features two collimated $^{55}$Fe sources (X-rays of 5.9$\,$keV and 6.5$\,$keV) for energy calibration, positioned to each shine on a side surface near one of the two sensors (see Figure \ref{double-picture}) to investigate any potential position dependence of the signal.

\begin{figure}[h!]
\begin{center}
\includegraphics[width=0.9\linewidth, keepaspectratio]{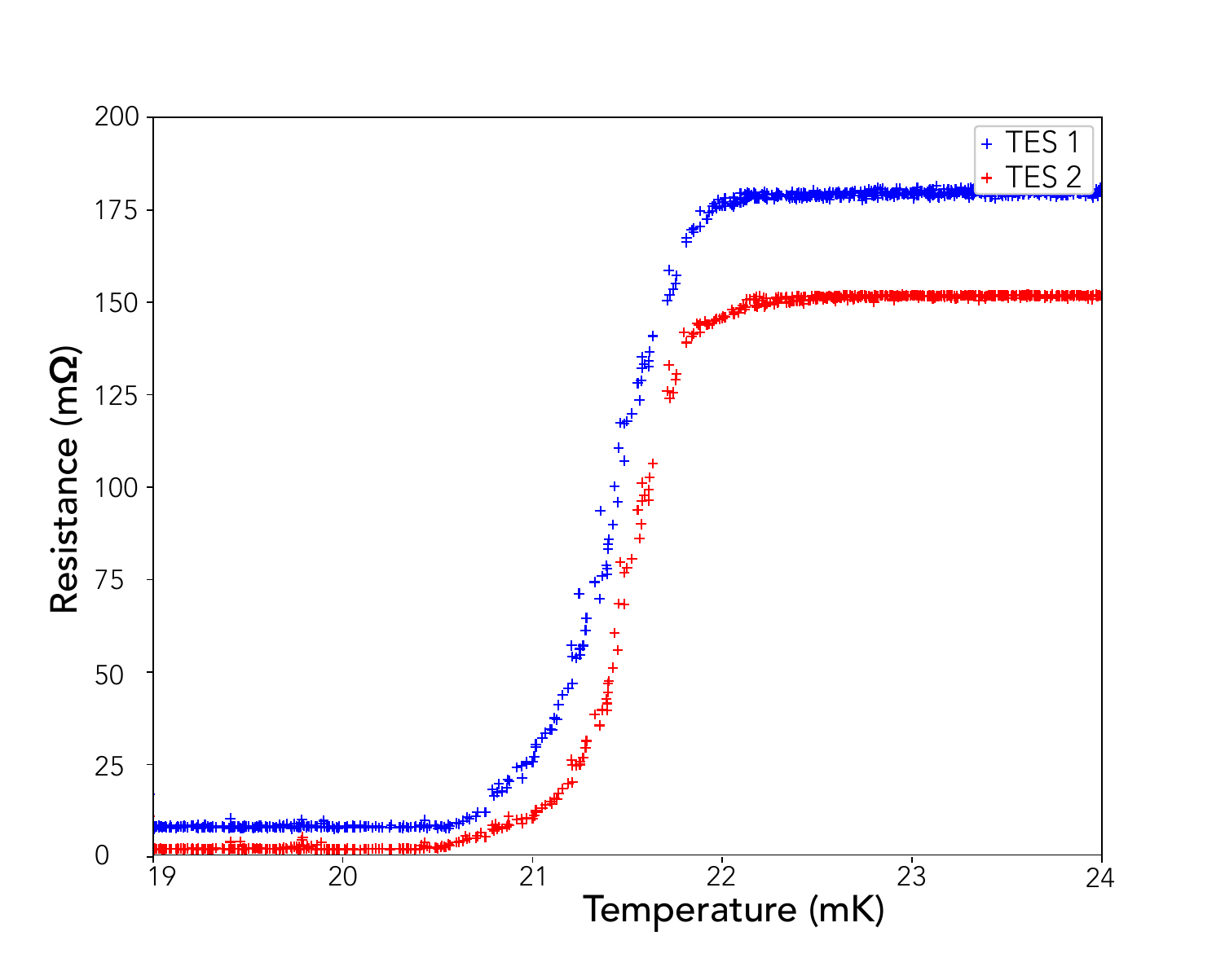}
\caption{Transition curves of the two TESs of the CaWO$_4$ doubleTES detector.}\label{transitions}
\end{center}
\end{figure}

In the initial phase of our measurement campaign, we focused on understanding the thermal interactions between the two sensors and their impact on performance optimisation. In the following, the two TESs will be referred to as "TES1" and "TES2".
Our study involves a thermal analysis, considering each sensor's transition temperature and heating requirements. Both sensors are equipped with heaters to maintain the TESs at the operating temperature and to inject heat pulses to monitor the detector response. Since both are on the same crystal, we anticipate some degree of thermal cross-talk. The issue arises particularly when the selected operating temperatures of the two sensors differ significantly. \\
Considering the thermal properties of the tested sensors and the temperature of the bath during operation, we can obtain an empirical condition for the module's operability, constraining the maximum allowable difference in transition temperatures between the two TESs for a given temperature of the thermal bath. In the configuration of the detector in use, this condition was fully satisfied with the two TESs having less than 0.5$\,$mK difference in T$_{\text{C}}$ (see Figure \ref{transitions}).

The second part of the measurement consisted in the performance studies of the two sensors. 
A total of 62 hours of measurement were collected.
The analysis procedure includes all the steps described in \ref{appendix:analysis}.

In this detector, the two TESs achieved a baseline resolution of (27.1$\pm$0.3)$\,$eV and (29.6$\pm$0.3)$\,$eV and a resolution at 5.9$\,$keV of (117$\pm$3)$\,$eV and (128$\pm$2)$\,$eV, respectively. The calibration of each single TES is obtained studying its energy spectrum independently. The calibration factor is obtained with a fit of the energy peaks provided by the $^{\text{55}}$Fe sources. The trigger thresholds for both sensors are set at five times the baseline resolution (5$\,$$\sigma_{\text{BL}})$, corresponding to 137$\,$eV for TES1 and 148$\,$eV for TES2. These thresholds are too high to study the full LEE spectrum as measured in the CRESST experiment but still allow to study of the presence of sensor-related events.

\begin{figure}[h!]
\begin{center}
\includegraphics[width=1.0\linewidth, keepaspectratio]{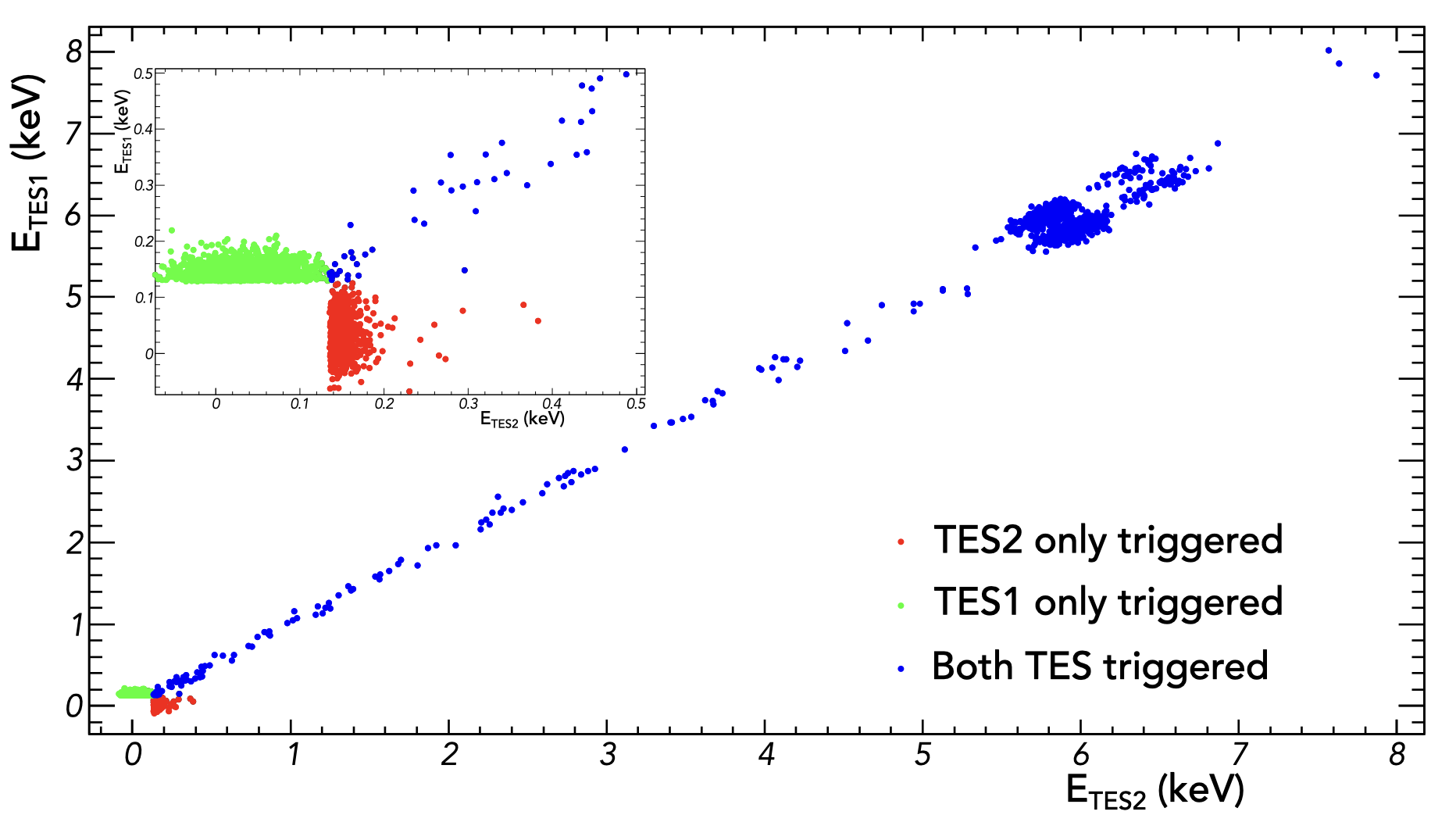}
\caption{2D plot with the energy of events in TES2 on the x-axis and the energy of events in TES1 on the y-axis. The three different populations of events are highlighted in different colours: in green, the events that triggered in TES1 only; in red, the events that triggered in TES2 only; and in blue, the events that triggered in both channels. In the inset, we provide a zoom of this plot into the energy region below 500$\,$eV.}\label{2Denergy_colours}
\end{center}
\end{figure}

After these performance studies on the single TESs, we performed a combined analysis of the two sensors to investigate the possibility of locating the origin of the detected events. The scatter plot presented in Figure \ref{2Denergy_colours} shows the energies measured event-by-event in the two TESs. Three different populations can be distinguished: a diagonal band representing the events measured with similar energy in both sensors and two populations at low energy in which only one TES recorded a pulse. Within the diagonal band, we observe two distinct $^{\text{55}}$Fe populations, each corresponding to one of the two sources. The separation between these populations is minimal, with the closer TES measuring at most 5$\%$ more energy than the distant TES, indicating only a slight position dependence in the detector response. To select only the events originating in the bulk of the absorber crystal, we conservatively require that the energies measured in coincidence by the two sensors do not differ by more than 35$\%$.

\begin{figure}[h!]
\begin{center}
\includegraphics[width=1\linewidth, keepaspectratio]{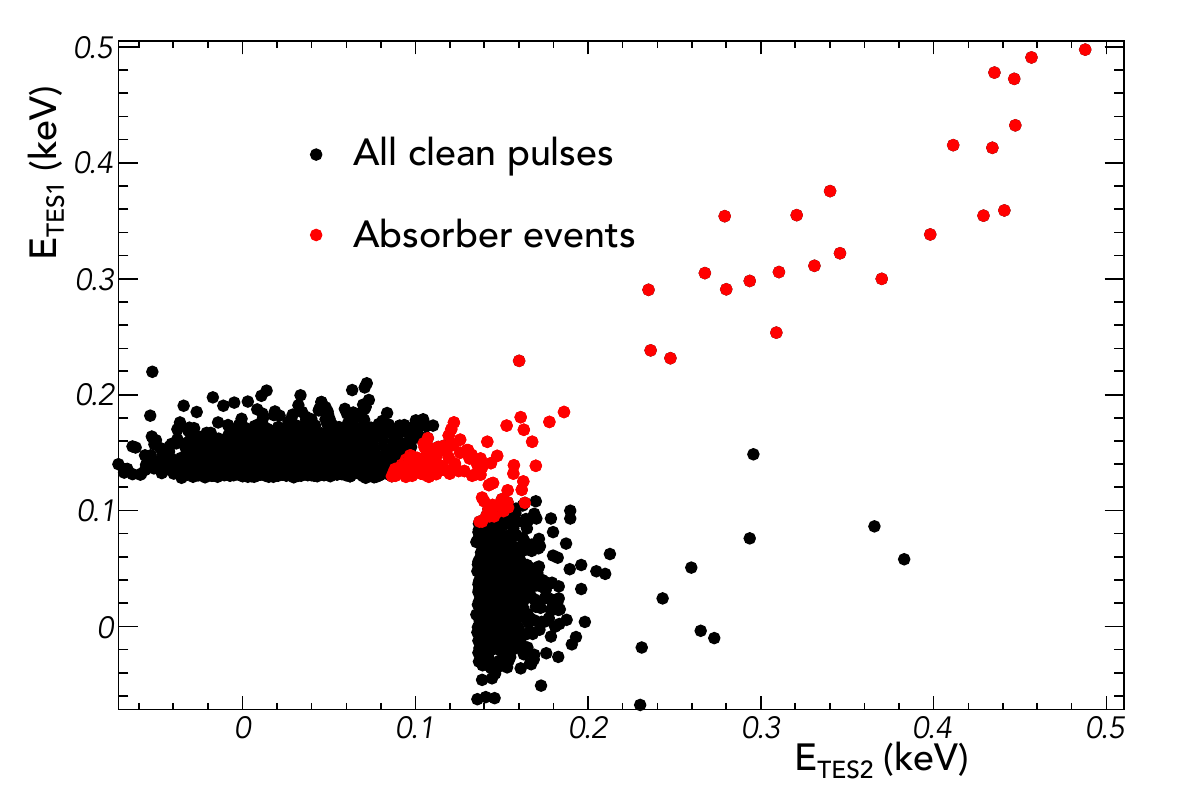} 
\caption{Zoom into the low energy region (up to 500$\,$eV) of the 2D plot with the energy deposited in TES2-L on the x-axis and the energy deposited in TES1-L on the y-axis. The events highlighted in red are the events accepted by the absorber events cut.}\label{CaWO4_2Dcut}
\end{center}
\end{figure}

\begin{figure}[h!]
\begin{center}
\includegraphics[width=1.0\linewidth, keepaspectratio]{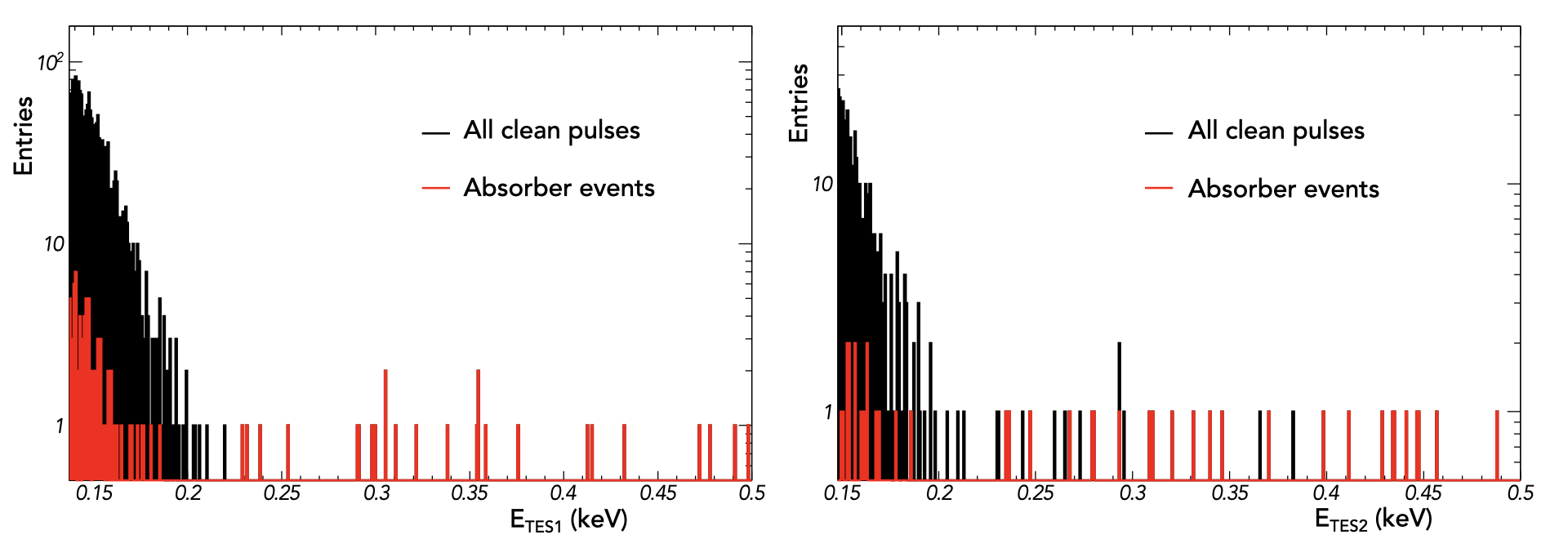} 
\caption{Energy spectra of TES1\,(left) and TES2\,(right) close to threshold. In black the spectra of the two TESs after quality cuts and in red the spectra after the quality cuts + absorber events cut.}\label{1D_spectra}
\end{center}
\end{figure}

\begin{figure}[h!]
\begin{center}
\includegraphics[width=1.0\linewidth, keepaspectratio]{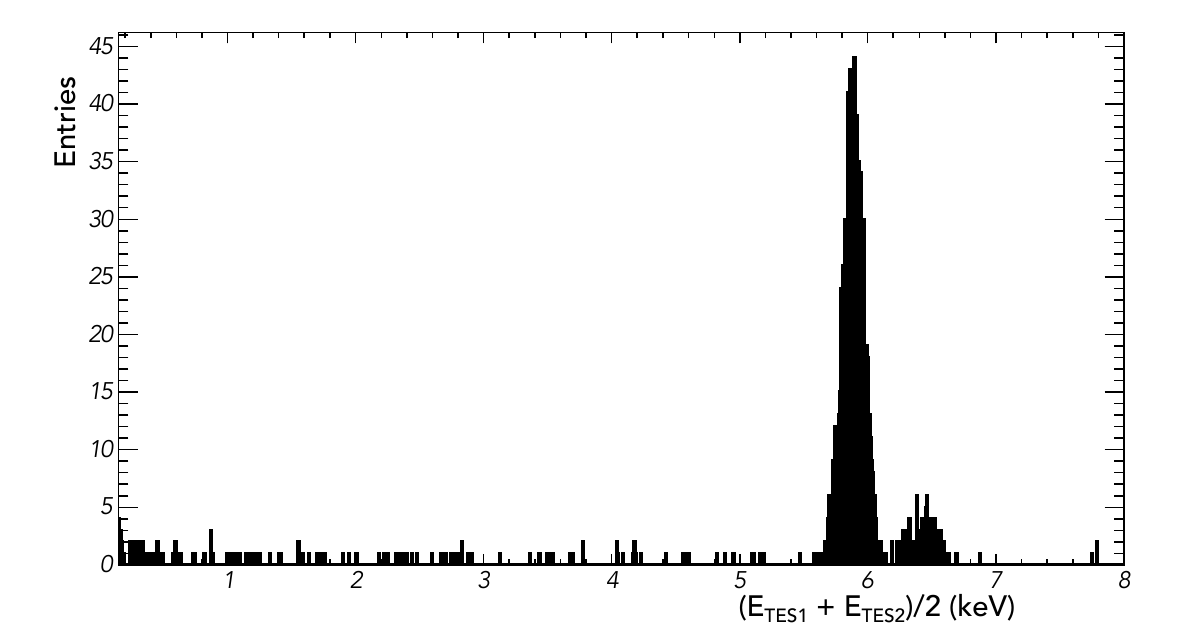} 
\caption{Energy spectrum of the averaged signal from the two TESs of the CaWO$_{\text{4}}$ detector. The two peaks from the 5.9$\,$keV and 6.5$\,$keV X-rays coming from the $^{55}$Fe calibration sources are well separated, showing an energy resolution of (81$\pm$2)$\,$eV at 5.9$\,$keV. }\label{spec_cawo}
\end{center}
\end{figure}

Implementing this absorber events cut significantly reduces the number of events close to threshold observed in the above-ground test of this module, as depicted in Figures \ref{CaWO4_2Dcut} and \ref{1D_spectra}. Given the observed position dependence, a more precise energy estimate is achieved through the averaging of the signal amplitude estimations from both sensors. The resulting spectrum, which combines the outputs from both TESs, is presented in Figure \ref{spec_cawo}, showing an improved energy resolution of (81$\pm$2)$\,$eV at 5.9$\,$keV.\\


The instabilities of the detector in the above-ground environment and the large dead time, mainly caused by pile-up, resulted in an 80$\%$ reduction of the collected statistics for both sensors.

Although the high thresholds of this detector combined with the above-ground environment of the test preclude any definitive conclusions about the LEE observed underground, we have measured events where a signal is recorded by only one of the sensors. These events are inherent to the sensors and are probably a constant presence in measurements. Calibrating such events is challenging because their response could differ significantly from that of energy depositions in the absorber.

\section{Silicon-on-sapphire doubleTES detector}\label{SOS}
To address the challenge posed by the high rate of events in the measurement with the CaWO$_4$ crystal and to access to lower energies, we developed and tested a doubleTES sensor on a smaller, 20$\times$20$\times$0.4$\,$mm$^3$, silicon-on-sapphire (SOS) crystal. The reduced size of this crystal is intended to achieve two key objectives: firstly, to decrease the overall rate of events, and secondly, to lower the detection threshold \cite{Threshold_mass}.\\
\begin{figure}[b]
\begin{center}
\includegraphics[width=0.8\linewidth, keepaspectratio]{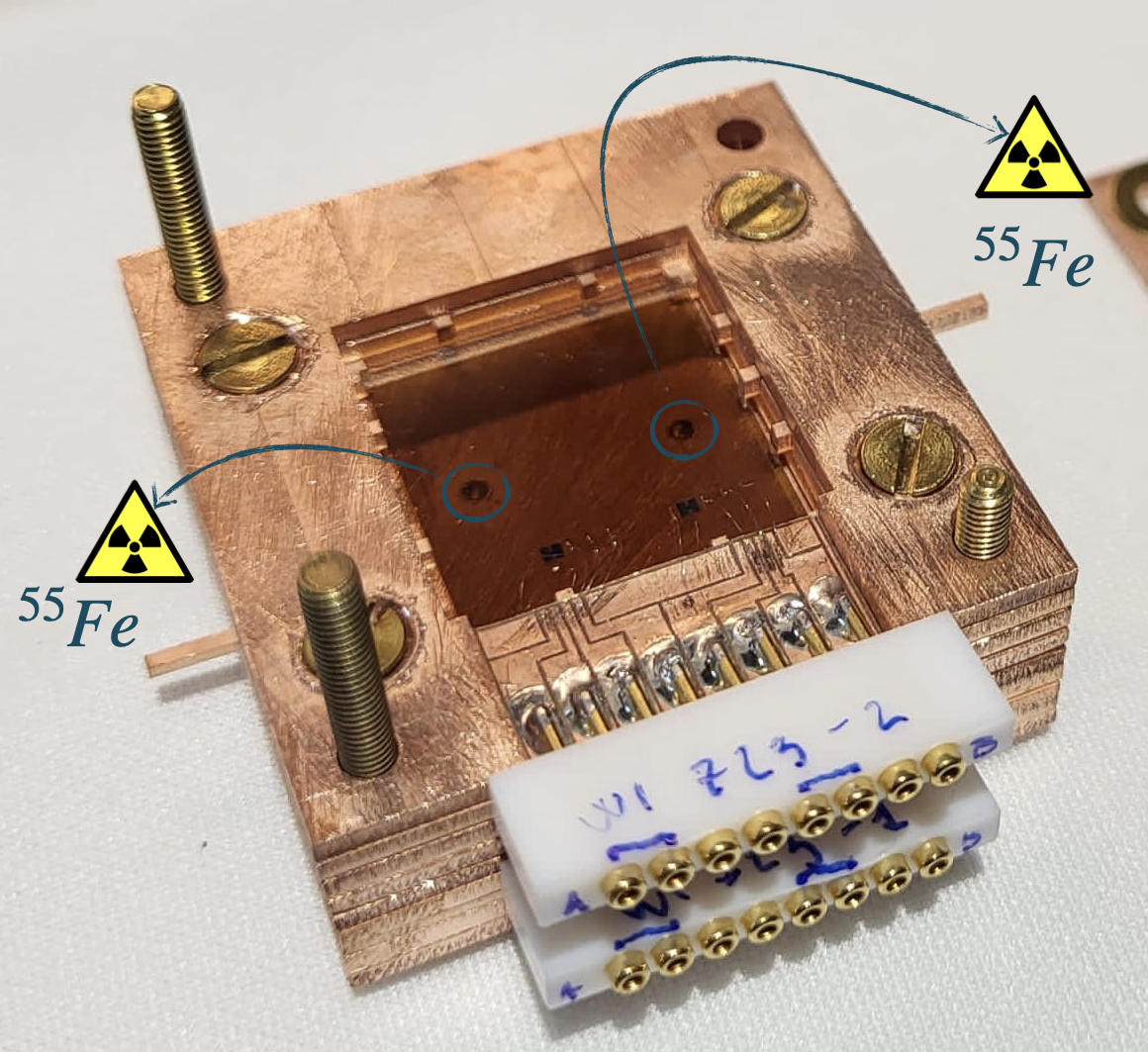} 
\caption{A picture of the doubleTES silicon-on-sapphire crystals, showing the two identical TESs. The two $^{\text{55}}$Fe sources are visible through the crystal and are circled in blue.}\label{SOS_foto}
\end{center}
\end{figure}

\subsubsection*{Test results}
\begin{figure}[h!]
\begin{center}
\includegraphics[width=1.0\linewidth, keepaspectratio]{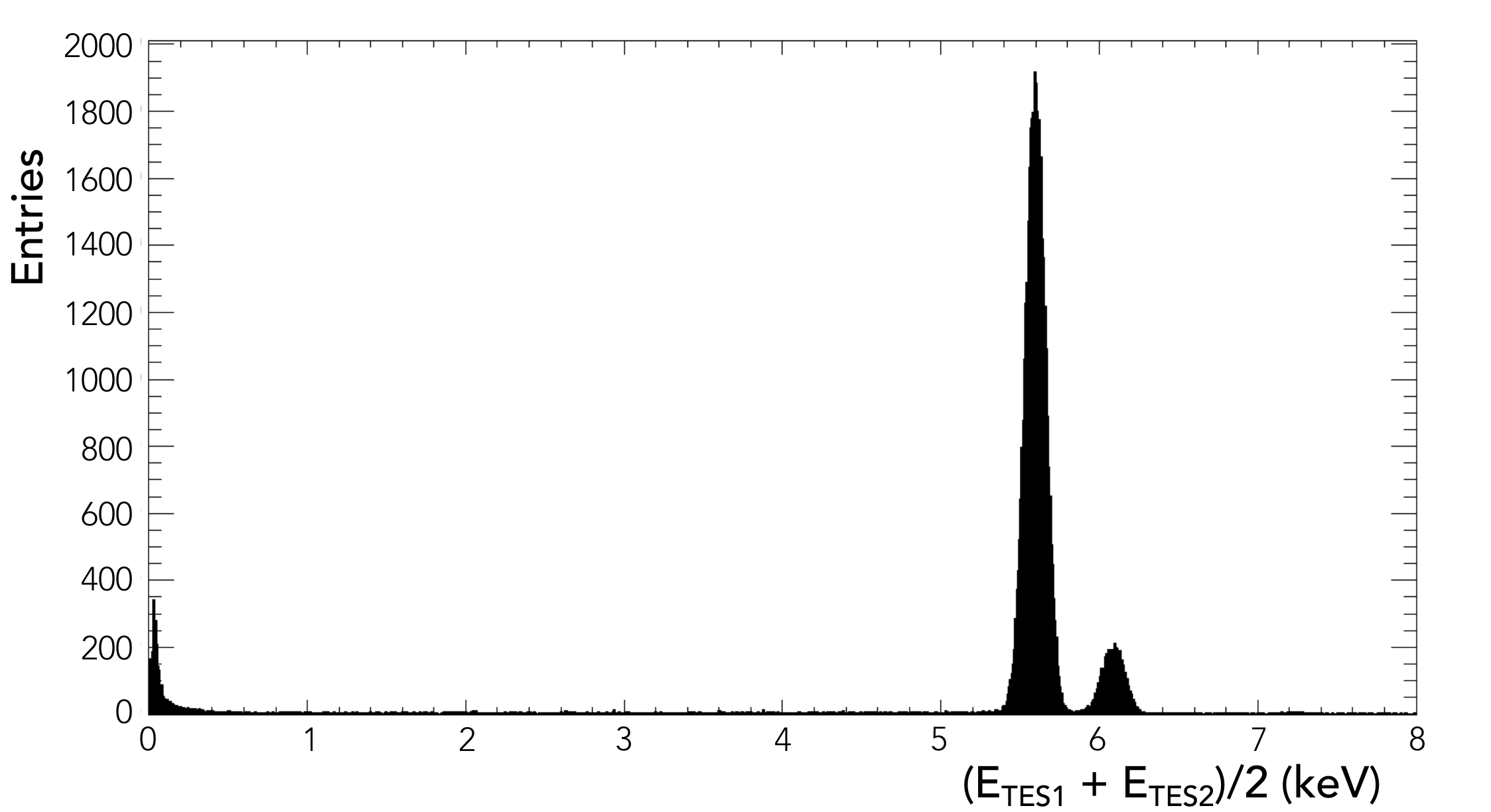} 
\caption{Averaged energy spectra of the two sensors of the SOS doubleTES. The two peaks from the 5.9$\,$keV and 6.5$\,$keV X-rays coming from the $^{55}$Fe calibration sources are very well separated, showing an energy resolution at 5.9$\,$keV of (90.0$\pm$1.1)$\,$eV.  }\label{SOS_Sum}
\end{center}
\end{figure}

The measurement was conducted using a dilution refrigerator at the Max-Planck Institute for Physics facility in Munich. The detector featured two collimated $^{55}$Fe sources for energy calibration, positioned symmetrically on the back side of the crystal, with the maximal possible separation to evaluate the position dependency of the detector response. The setup is shown in Figure \ref{SOS_foto}.
The analysis followed the same procedures outlined in \ref{appendix:analysis}. In this module, the two TESs, named TES1-L and TES2-L, exhibited outstanding performance. TES1-L achieved a baseline resolution of (5.4$\pm$0.1)$\,$eV and a resolution of (149.0$\pm$3.8)$\,$eV at 5.9$\,$keV, while TES2-L reached (4.1$\pm$0.1)$\,$eV and (121.6$\pm$2.3)$\,$eV, respectively. We set the trigger threshold at five times the baseline resolution, corresponding to 27$\,$eV for TES1-L and 20.5$\,$eV for TES2-L. By averaging the energy outputs of TES1-L and TES2-L, this module reaches an energy resolution of (90.0$\pm$1.1)$\,$eV at 5.9$\,$keV (see Figure \ref{SOS_Sum}). \\

\begin{figure}[t]
\begin{center}
\includegraphics[width=1\linewidth, keepaspectratio]{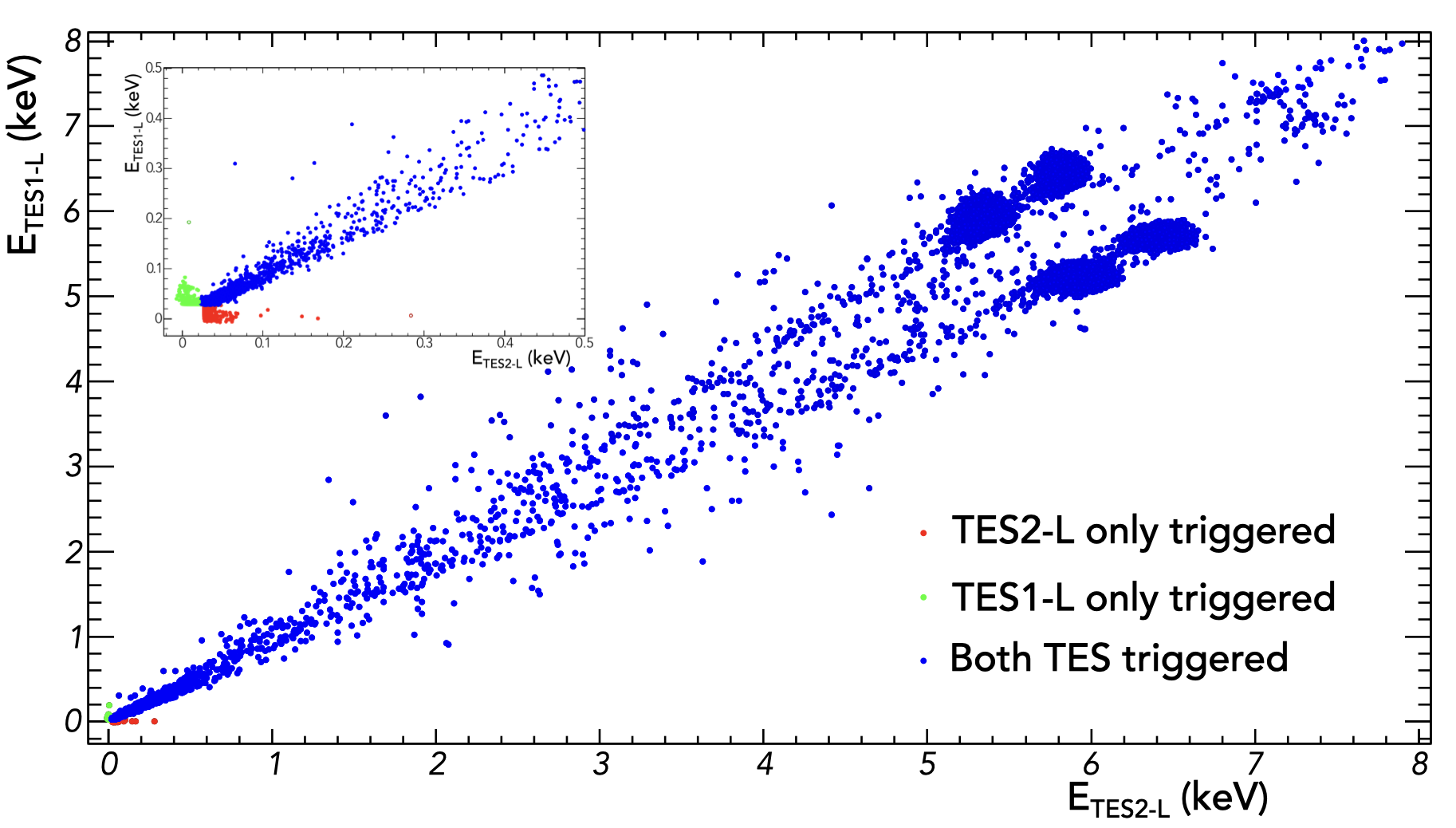} 
\caption{2D plot with the energy deposited in TES2-L on the x-axis and the energy deposited in TES1-L on the y-axis. The three different populations of events are highlighted in different colours: in green, the events that triggered in TES1-L only; in red, the events that triggered in TES2-L only; and in blue, the events that triggered in both channels. In the inset, we provide a zoom into the energy region below 500$\,$eV.}\label{SOS_2D}
\end{center}
\end{figure}

\begin{figure}[t]
\begin{center}
\includegraphics[width=1\linewidth, keepaspectratio]{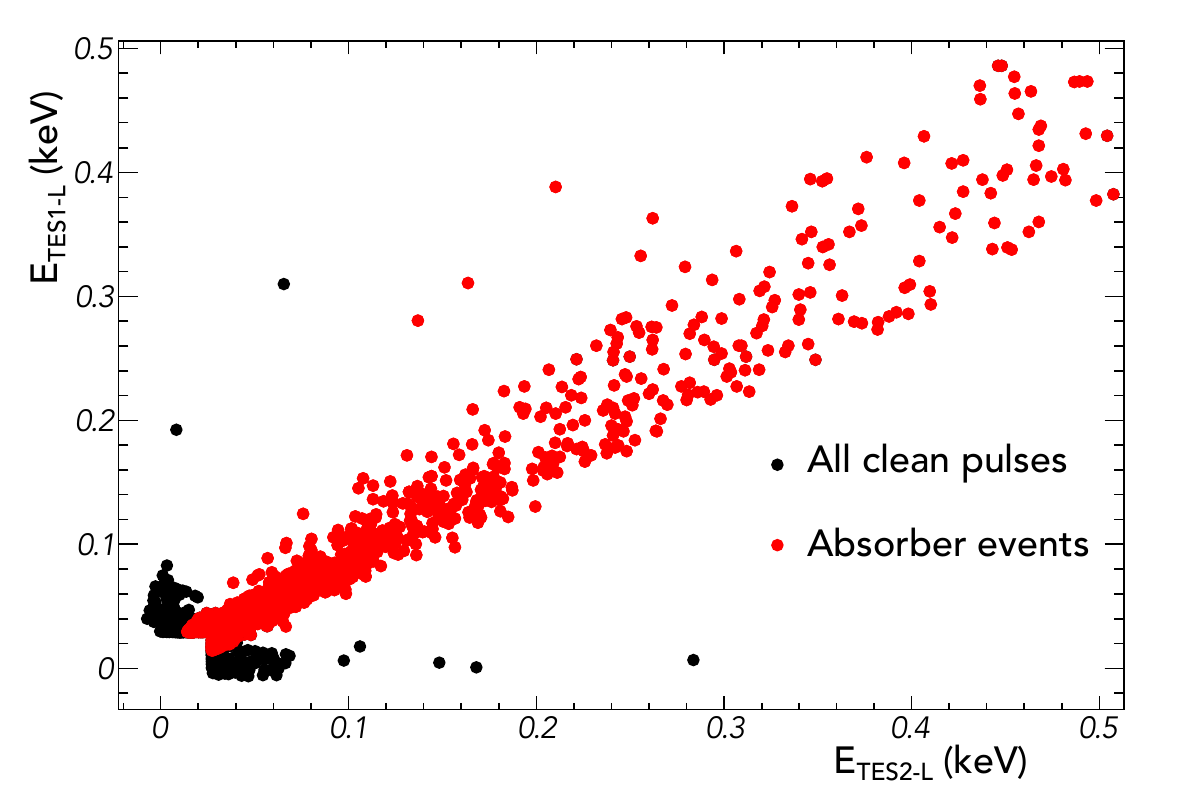} 
\caption{Zoom into the low energy region (up to 500$\,$eV) of the 2D plot with the energy deposited in TES2-L on the x-axis and the energy deposited in TES1-L on the y-axis. The events highlighted in red are the events accepted by the absorber events cut.}\label{SOS_2Dcut}
\end{center}
\end{figure}

The 2D scatter plot in Figure \ref{SOS_2D} shows three distinct populations: one per single TES and a population with approximately the same energy in the two sensors. Compared to the previously discussed CaWO$_4$ crystal, we observed a more significant position-dependent variation in the energy sharing of the two individual TESs, attributable to the flat geometry of the target crystal. In each of the two $^{\text{55}}$Fe populations, the TES closer to the source measures up to 15$\%$ more energy than the distant TES. To account for the stronger position dependence observed in this dataset, we classified events as absorber events when the energy difference of the signals in the two sensors is less than 50$\%$.
This measurement clearly shows that the LEE is composed of multiple components.\\
Like in the previous test, one component includes events that occur close to one of the sensors. Similar to the first measurement with CaWO$_4$ (see Section \ref{CaWO4}), such events occurring near the sensors are not expected to align with the calibration of events within the bulk of the crystal.\\
In the absorber events band, we observe a significant rise in the event rate below 150$\,$eV. The 2D scatter plot illustrating these sub-300$\,$eV events is presented in Figure \ref{SOS_2Dcut}.  
\\

\begin{figure}[htbp]
\begin{center}
\includegraphics[width=1.0\linewidth, keepaspectratio]{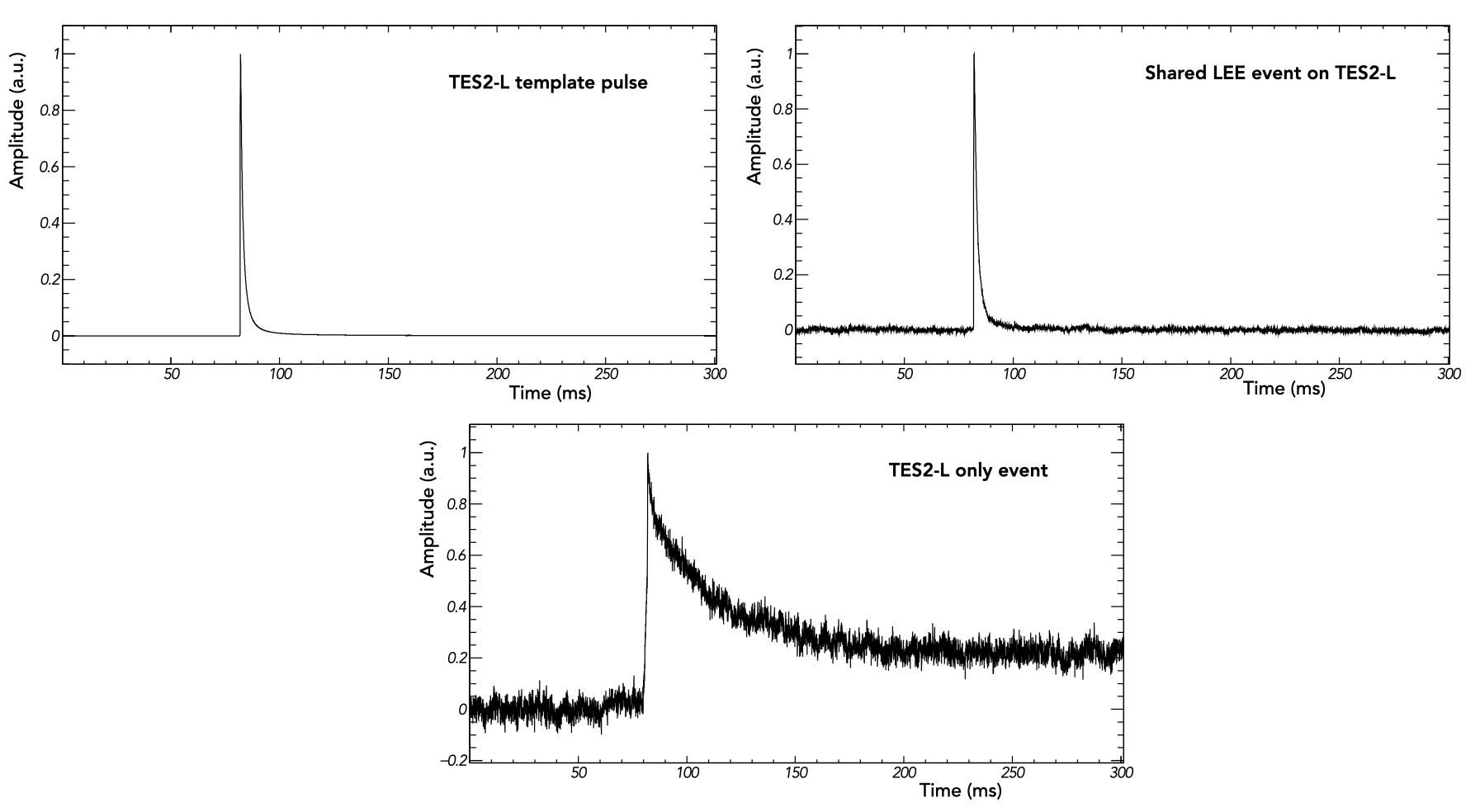} 
\caption{Comparison of different pulses observed by TES2-L. The top left plot corresponds to the template pulse obtained averaging some pulses from the 5.9$\,$keV X-ray peak from the $^{55}$Fe source. The top right plot displays an average pulse from the diagonal band at very low energy, obtained by averaging pulses with energies between 30$\,$eV and 50$\,$eV. The bottom plot represents an average pulse of the "singles" event band, obtained by averaging pulses with energies above 40$\,$eV. }\label{Pulses_Comparisons}
\end{center}
\end{figure}

Ensuring that the observed events at low energy are not noise events is crucial. To address this concern, we studied the spectra extracted from the inverted data stream. These tests, analogously to what shown in \cite{siliconpapero}, showed that the events constituting the LEE occur only with a positive trigger, while noise triggers would be expected with both polarities. Moreover we averaged pulses from different populations. Figure \ref{Pulses_Comparisons} illustrates the resulting pulses alongside the template pulses (see Section \ref{appendix:analysis}) of this dataset obtained by averaging pulses from the calibration source peaks. Notably, all populations consist of genuine pulses. As expected, in the diagonal band, the pulses exhibit the same pulse shape as the template pulses. The off-diagonal components show a different pulse shape. A different pulse shape between events in the diagonal band and events in the single TES only was also observed in the data set described in Section \ref{CaWO4}, although the differences between the two datasets show opposing patterns (slower single TES pulses here and faster single TES pulses for the data set presented in Section \ref{CaWO4}). Despite the distinctive pulse shapes, the single TES pulses cannot be effectively discriminated at energies close to threshold. The unique information provided by the doubleTES design is essential for accurately isolating these pulses and removing this component of the LEE.

\subsubsection*{Time dependence studies}
\begin{figure}[!]
\begin{center}
\includegraphics[width=1\linewidth, keepaspectratio]{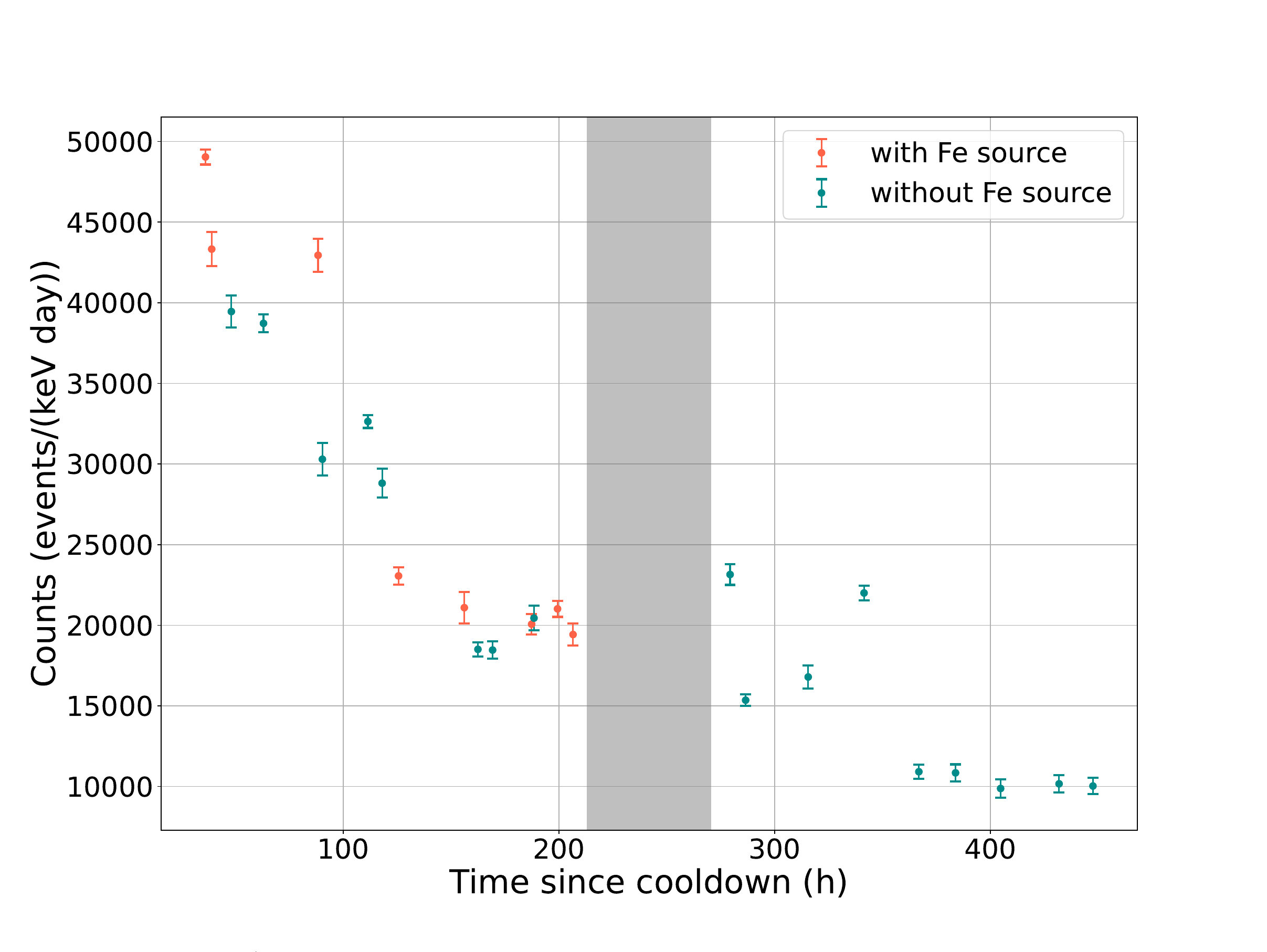} 
\includegraphics[width=1\linewidth, keepaspectratio]{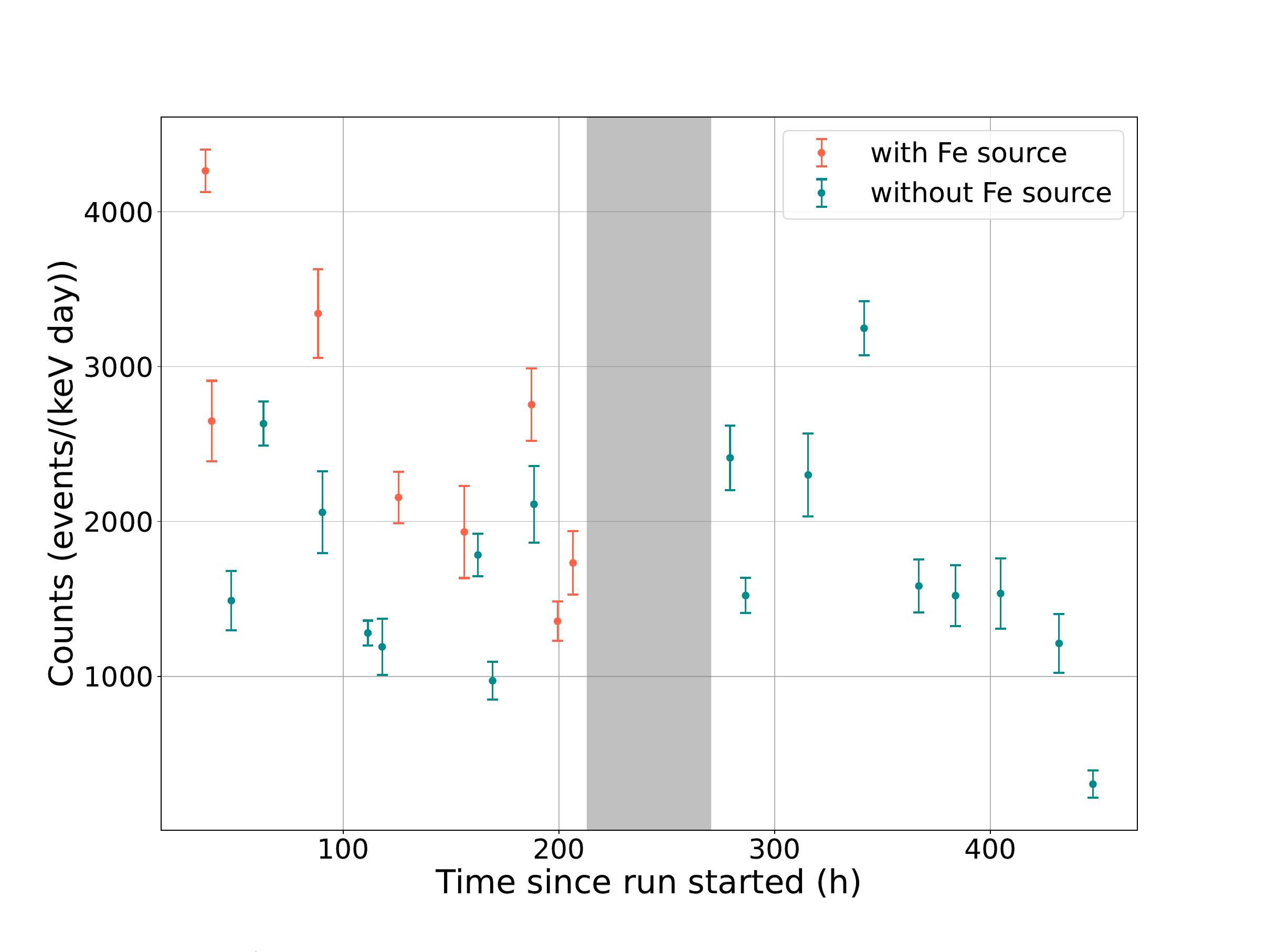}
\caption{\textit{Top:} Time evolution for the absorber events band of the above-ground LEE for events with energies between 28$\,$eV and 50$\,$eV in the two experimental setups tested. The data points from the two runs appear compatible with the same decay. The decay times are obtained with single exponential fits. The data with the $^{\text{55}}$Fe source shows a decay time of (7.4$\pm$1.1)$\,$days, while the data without the calibration source show a decay time of (12.4$\pm$1.5)$\,$days. The decay time computed using the data points from both sets is (10.2$\pm$1.1)$\,$days. \textit{Bottom:} Time evolution of the TES2-L only event rate of the above-ground LEE for events with energies above 40$\,$eV combined for the two experimental setups. For both plots, the red dots correspond to the rates from the measurement with the $^{\text{55}}$Fe sources, and in blue to the rates from the measurement without it. Every point of these plots corresponds to a single data file, and the statistical errors are computed for each file separately. The rates are corrected for the calculated survival probability (probability that a pulse originating from
a particle interaction contributes to the final spectrum after analysis, see \ref{appendix:analysis}), computed file by file. The regions highlighted in grey show the time when the lead shield was built for the measurement without the $^{\text{55}}$Fe during which no data was collected.}\label{time_evo_all}
\end{center}
\end{figure}

To verify that our observation is consistent with the LEE measured by CRESST detectors, we studied the time dependence of the rate of the low energy events.\\
Given the consistently higher rates of excess events observed in above-ground setups compared to underground measurements (with above-ground rates typically being at least 1-2 orders of magnitude higher), we were prompted to investigate the influence of external radiation. This factor stands out as a key difference between the above-ground setups and the underground facility used by CRESST. To address this, we carried out a second measurement without the $^{\text{55}}$Fe sources. This measurement was crucial to eliminate any potential contributions from our calibration source, which previously accounted for half of the total event rate. Minimal alterations were made to the setup; only the sources were removed, while the cabling and the settings for bias and heater currents remained unchanged. Without a calibration source, for the second measurement we used the calibrated response to the heat pulses from the first measurement.\\
Figure \ref{time_evo_all} (\textit{top}) shows that both the event rate of the diagonal component of the LEE and its time evolution are similar with and without the $^{\text{55}}$Fe calibration source. The rate of absorber events with energies between 28$\,$eV and 50$\,$eV decreases with a decay time of (10.2 $\pm$ 1.1)\,days, obtained with a single exponential fit of the combined data set. This value is compatible with the average decay time of (18$\pm$7$\,$)\,days observed in the main CRESST setup for the fast component of the LEE \cite{LEEdescr}.
Form these observations we can conclude that the presence of the $^{\text{55}}$Fe source does not significantly contribute to the above-ground LEE in the absorber events band,  consistently with what was observed in the underground measurements.\\
To further support this finding and study the impact of external radiation, we constructed a 10~cm thick lead shield around the setup roughly 200 hours after the beginning of the run without calibration source. While the presence of the lead shield substantially changed the background conditions, with an overall reduction of the event rate by a factor of three for energies above the endpoint of the LEE ($\approx$220eV), this had no significant impact on the above-ground LEE in the absorber events band.

The similarity of the time behaviour observed in this measurement campaign with the one observed in underground measurements \cite{LEEdescr} could indicate that the LEE observed share the same origin. 
The difference in the observed event rates and in the time averaged spectra could be attributed to the fact that in above-ground measurements, data-taking usually commences only a few hours after cooldown and concludes within a few days, whereas in underground measurements, data-taking typically initiates some days after cooldown and spans over several months. Consequently, the differences might be partially attributed to the observation of the decay components at different times after the cooldown. Further tests are needed to verify the compatibility of above-ground and underground measurements.\\
We also examined the rates of events appearing in one W-TES only, focusing on events above 40$\,$eV to exclude leakage of events from the absorber band. The variation of these rates over time, from the start of the run, is illustrated in Figure \ref{time_evo_all} (\textit{bottom}) for setups with and without the $^{\text{55}}$Fe sources. While for the measurement without calibration source the data points do not show a decay pattern, the data with the calibration source slightly prefer a decreasing rate, although within the very limited time frame of the measurement. The tension between the observations indicates the need for additional long measurements. In case of a constant or slowly decaying rate, these off-diagonal events could become a dominant contribution to the low energy excess over a long data-taking period. Such a component of the LEE could be efficiently tagged using the information from the double sensor readout, potentially increasing the sensitivity to DM interactions in the CRESST experiment.

\section{Conclusions}
This work details the findings from above-ground tests on two crystals using a detector design with a double sensor readout. A key observation across all measurements is the detection of multiple low energy excess components. We observe the component that generates a signal in only one of the two TESs at the energy threshold for all measurements presented in this work. These events likely do not share the energy calibration of bulk events in the absorber crystal.\\
With a low threshold doubleTES, we looked into the energy spectrum below 150$\,$eV, identifying an absorber events component of the low energy excess observed above-ground that produces similar signals in both sensors. This component decreases exponentially over time, with a time constant of (10.2$\pm$1.1)$\,$days.\\
The events of both components are not compatible with misidentified noise, confirmed both by noise polarity tests and by studying the averaged pulses from different populations.\\
We also ruled out a major contribution from external radiation or the calibration source to the observed low energy spectrum.\\
Studying the time behaviour of the LEE in the above-ground measurements, we observe decay times compatible with the underground observations.\\
These findings might indicate that the LEE events observed in above-ground measurements have a similar origin to those observed in the CRESST underground setup. If this is the case, the difference in event rates inferred from the datasets acquired in the different setups could be attributed to some extent to the faster starts and shorter duration of above-ground experiments. This insight is crucial for a consistent interpretation of data across various experimental conditions. Nevertheless, further above-ground and underground tests are planned for a comprehensive investigation of this possibility.\\
The doubleTES design proved to be able to distinguish different components of the event excess at threshold. This distinction is critical in our ongoing quest to understand the origins of the LEE. Future underground measurements at the CRESST facility are expected to provide further insights into this background, which currently poses a challenge to the direct detection of dark matter, especially in the low-mass range.\\
The advancements and findings from this study contribute to a deeper understanding of the LEE and pave the way for more sensitive dark matter searches.

\begin{acknowledgements}

We are thankful to the COSINUS MPP group for promptly providing us with 10$\,$cm thick lead bricks for the shielding, fundamental for the studies of this paper. We are grateful to Laboratori Nazionali del Gran Sasso - INFN for their generous support of CRESST. This work has been funded by the Deutsche Forschungsgemeinschaft (DFG, German Research Foundation) under Germany’s Excellence Strategy – EXC 2094 – 390783311 and through the Sonderforschungsbereich (Collaborative Research Center) SFB1258 ‘Neutrinos and Dark Matter in Astro- and Particle Physics’, by the BMBF 05A20WO1 and 05A20VTA and by the Austrian Science Fund (FWF): I5420-N, W1252-N27 and FG1 and by the Austrian research promotion agency (FFG), project ML4CPD. JB and HK were funded through the FWF project P 34778-N ELOISE. The Bratislava group acknowledges the support provided by the Slovak Research and Development Agency (projects APVV-15-0576 and APVV-21-0377).
\end{acknowledgements}

\bibliographystyle{modified_spphys}
\bibliography{main.bib}

\appendix
\section{Data Analysis}\label{appendix:analysis}
In the following, we describe the analysis procedure of the two above-ground tests conducted on the doubleTES modules. The data are recorded continuously, with a sampling frequency of 25$\,$kHz.
\begin{enumerate}
\item \textbf{Trigger}: we initiate the trigger process by constructing a matched filter using the distinctive noise power spectrum and template pulse characteristics of a detector \cite{Gatti:OFilter}. The template pulse is a template of the known pulse shape of interest obtained by averaging a set of particle pulses of the TES. The filter is then applied to the data stream. The filtered stream undergoes a triggering algorithm to detect pulses. Notably, the acquisition software supplies us with information regarding the time stamps of test pulses, enabling us to precisely tag and identify these specific pulses within the dataset. \\
The test pulses are electrical pulses injected into the detector through the heater element. They vary in amplitude, covering the entire energetic dynamic range of the detector. Test pulses serve two primary purposes: firstly, to verify the stability of the detector's response, and secondly, to extend the energy calibration across the full dynamic range of the detector.
\item \textbf{Pulse Parameter Calculation}: After identifying pulses through the trigger, we compute essential parameters such as amplitude, rise time, decay time, and others, providing a detailed characterisation of each pulse.
\item \textbf{Stability Cut}: the resistance value of the sensors during normal operation describes their operating point. Using the heater, we maintain the detector at a specific temperature to stabilise it at its optimal operating point, resulting in a particular resistance. We exclusively consider pulses originating from specific resistance values around the chosen working point for both TESs.
\item \textbf{Quality Cuts}:  we discard pulses with issues such as tilted baselines, resetting pulses, and electronic artefacts by applying basic quality cuts on pulse parameters.
\item \textbf{Calibration}: using the $^{\text{55}}$Fe calibration source emitting X-rays at 5.9$\,$keV and 6.5$\,$keV, alongside test pulses, we conduct a two-step calibration process. Initially, we cross-calibrate the test pulses from the heater with the 5.9$\,$keV peak from the $^{\text{55}}$Fe source. Subsequently, this calibration is extended across the entire energy dynamic range, employing the presence of multiple test pulses. This procedure involves an event-by-event based calibration, where every event is calibrated taking into account the working point of the detector at the specific time of the event.
\item \textbf{Absorber Events Cut}: this cut has been specifically designed for this module as a first idea on how to discriminate events with different origins. As described in Section \ref{sec:doubleTES_motivation}, we expect the two TESs to measure roughly the same energy after a particle interaction in the bulk of the crystal, whereas interactions occurring in close proximity to one of the two TESs are expected to yield different energy measurements for each. Hence, we design the absorber events cut based on the ratio of energies measured by the two TESs. We select events where the ratio of the two energies is within a certain percentage $x$ around 1 as absorber or bulk events:
\begin{equation}
\begin{cases}
    \frac{E_{1}}{E_{2}} > 1 -x/100 \\
    \frac{E_{2}}{E_{1}} > 1 -x/100 
\end{cases}
\end{equation}
The value $x$ depends on the performance of the module and is adjusted for each measurement.  This cut is designed differently for each measurement. 
\item \textbf{Survival Probability Calculation}: calculating efficiency consists in assessing the probability that a pulse originating from a particle interaction contributes to the final spectrum. This is achieved by simulating events with the signal pulse shape in both channel simultaneously across our data stream, conducting an identical analysis to that performed on the raw stream, and calculating the fraction of the remaining simulated events.
\end{enumerate}

\end{document}

%% file: authors_cresst_EPJ_format.tex
\author{
  G.~Angloher\thanksref{addrMPI}\and
  S.~Banik\thanksref{addrHEPHY,addrAI}\and
  G.~Benato\thanksref{addrLNGS,addrGSSI}\and
  A.~Bento\thanksref{addrMPI,addrCoimbra}\and
  A.~Bertolini\thanksref{addrMPI}\and 
  R.~Breier\thanksref{addrBratislava}\and
  C.~Bucci\thanksref{addrLNGS}\and 
  J.~Burkhart\thanksref{addrHEPHY}\and 
  L.~Canonica\thanksref{addrMPI,addrMIL}\and 
  A.~D'Addabbo\thanksref{addrLNGS}\and
  S.~Di~Lorenzo\thanksref{addrMPI}\and
  L.~Einfalt\thanksref{addrHEPHY,addrAI}\and
  A.~Erb\thanksref{addrTUM,addrWMI}\and
  F.~v.~Feilitzsch\thanksref{addrTUM}\and 
  S.~Fichtinger\thanksref{addrHEPHY}\and
  D.~Fuchs\thanksref{addrMPI}\and 
  A.~Garai\thanksref{addrMPI}\and 
  V.M.~Ghete\thanksref{addrHEPHY}\and
  P.~Gorla\thanksref{addrLNGS}\and
  P.V.~Guillaumon\thanksref{addrMPI,addrSAOP}\and
  S.~Gupta\thanksref{addrHEPHY}\and 
  D.~Hauff\thanksref{addrMPI}\and 
  M.~Ješkovsk\'y\thanksref{addrBratislava}\and
  J.~Jochum\thanksref{addrTUE}\and
  M.~Kaznacheeva\thanksref{addrTUM}\and
  A.~Kinast\thanksref{addrTUM}\and
  H.~Kluck\thanksref{addrHEPHY}\and
  H.~Kraus\thanksref{addrOxford}\and 
  S.~Kuckuk\thanksref{addrTUE}\and 
  A.~Langenk\"amper\thanksref{addrMPI}\and 
  M.~Mancuso\thanksref{addrMPI,t1}\and
  L.~Marini\thanksref{addrLNGS}\and 
  B.~Mauri\thanksref{addrMPI}\and 
  L.~Meyer\thanksref{addrTUE}\and 
  V.~Mokina\thanksref{addrHEPHY}\and
  M.~Olmi\thanksref{addrLNGS}\and
  T.~Ortmann\thanksref{addrTUM}\and
  C.~Pagliarone\thanksref{addrLNGS,addrCASS}\and
  L.~Pattavina\thanksref{addrLNGS,addrMIL2}\and
  F.~Petricca\thanksref{addrMPI}\and 
  W.~Potzel\thanksref{addrTUM}\and 
  P.~Povinec\thanksref{addrBratislava}\and
  F.~Pr\"obst\thanksref{addrMPI}\and
  F.~Pucci\thanksref{addrMPI,addrTUM,t2}\and 
  F.~Reindl\thanksref{addrHEPHY,addrAI} \and
  J.~Rothe\thanksref{addrTUM}\and 
  K.~Sch\"affner\thanksref{addrMPI}\and 
  J.~Schieck\thanksref{addrHEPHY,addrAI}\and 
  S.~Sch\"onert\thanksref{addrTUM}\and 
  C.~Schwertner\thanksref{addrHEPHY,addrAI}\and
  M.~Stahlberg\thanksref{addrMPI}\and 
  L.~Stodolsky\thanksref{addrMPI}\and 
  C.~Strandhagen\thanksref{addrTUE}\and
  R.~Strauss\thanksref{addrTUM}\and
  I.~Usherov\thanksref{addrTUE}\and
  F.~Wagner\thanksref{addrHEPHY,addrETH,addrPSI}\and 
  V.~Wagner\thanksref{addrTUM}\and
  V.~Zema\thanksref{addrMPI}
(CRESST Collaboration)
}

\institute
{Max-Planck-Institut f\"ur Physik, D-85748 Garching bei M\"unchen, Germany \label{addrMPI} \and
Institut f\"ur Hochenergiephysik der \"Osterreichischen Akademie der Wissenschaften, A-1050 Wien, Austria\label{addrHEPHY} \and
Atominstitut, Technische Universit\"at Wien, A-1020 Wien, Austria \label{addrAI} \and
INFN, Laboratori Nazionali del Gran Sasso, I-67100 Assergi, Italy \label{addrLNGS} \and
Comenius University, Faculty of Mathematics, Physics and Informatics, 84248 Bratislava, Slovakia \label{addrBratislava} \and
Physik-Department, TUM School of Natural Sciences, Technische Universität München, D-85747 Garching, Germany \label{addrTUM} \and
Eberhard-Karls-Universit\"at T\"ubingen, D-72076 T\"ubingen, Germany \label{addrTUE} \and
Department of Physics, University of Oxford, Oxford OX1 3RH, United Kingdom \label{addrOxford} \and
also at: LIBPhys-UC, Departamento de Fisica, Universidade de Coimbra, P3004 516 Coimbra, Portugal \label{addrCoimbra} \and
also at: Walther-Mei\ss ner-Institut f\"ur Tieftemperaturforschung, D-85748 Garching, Germany \label{addrWMI} \and
also at: GSSI-Gran Sasso Science Institute, I-67100 L'Aquila, Italy \label{addrGSSI} \and
also at: Dipartimento di Ingegneria Civile e Meccanica, Università degli Studi di Cassino e del Lazio Meridionale, I-03043 Cassino, Italy\label{addrCASS} \and
also at: Dipartimento di Fisica, Università di Milano Bicocca, Milano, 20126, Italy\label{addrMIL2} \and
also at: Instituto de Física, Universidade de São Paulo, São Paulo 05508-090, Brazil \label{addrSAOP}
\and
now at: Dipartimento di Fisica, Università di Milano Bicocca, Milano, 20126, Italy\label{addrMIL} \and
now at: Department of Physics, ETH Zurich, CH-8093 Zurich, Switzerland\label{addrETH} \and
now at:  ETH Zurich - PSI Quantum Computing Hub, Paul Scherrer Institute, CH-5232 Villigen, Switzerland\label{addrPSI}
}

\thankstext[$\dagger$]{t1}{Corresponding author: mancuso@mpp.mpg.de}
\thankstext[$\star$]{t2}{Corresponding author: frapucci@mpp.mpg.de}